\documentclass[a4paper]{jpconf}
\usepackage{amssymb,graphicx}
\begin{document}
\title{Quantum inflaton, primordial metric perturbations and CMB
fluctuations}

\author{F J Cao$^{1,2}$}

\address{$^1$ Departamento de F\'\i sica At\'omica, Molecular y
Nuclear, Universidad Complutense de Madrid, Avenida Complutense
s/n, 28040 Madrid, Spain}

\address{$^2$ LERMA, Observatoire de
Paris, Laboratoire Associ\'e au CNRS UMR 8112, 61, Avenue de
l'Observatoire, 75014 Paris, France}

\ead{francao@fis.ucm.es}

\begin{abstract}
We compute the primordial scalar, vector and tensor metric
perturbations arising from quantum field inflation. Quantum field
inflation takes into account the nonperturbative quantum dynamics
of the inflaton consistently coupled to the dynamics of the
(classical) cosmological metric. For chaotic inflation, the
quantum treatment avoids the unnatural requirements of an initial
state with all the energy in the zero mode. For new inflation it
allows a consistent treatment of the explosive particle production
due to spinodal instabilities. Quantum field inflation (under
conditions that are the quantum analog of slow roll) leads, upon
evolution, to the formation of a condensate starting a regime of
effective classical inflation. We compute the primordial
perturbations taking the dominant quantum effects into account.
The results for the scalar, vector and tensor primordial
perturbations are expressed in terms of the classical inflation
results. For a N-component field in a $ O(N) $ symmetric model,
adiabatic fluctuations dominate while isocurvature or entropy
fluctuations are negligible. The results agree with the current
WMAP observations and predict corrections to the power spectrum in
classical inflation. Such corrections are estimated to be of the
order of $ m^2/[N H^2] $ where m is the inflaton mass and H the
Hubble constant at horizon crossing. This turns to be about $ 4\%
$ for the cosmologically relevant scales. This quantum field
treatment of inflation provides the foundations to the classical
inflation and permits to compute quantum corrections to it.
\end{abstract}

\newcommand{\be}{\begin{equation}}
\newcommand{\ee}{\end{equation}}
\newcommand{\bea}{\begin{eqnarray}}
\newcommand{\eea}{\end{eqnarray}}

\section{Introduction}

Inflation is a stage of accelerated expansion in the very early
Universe \cite{infbooks,revkolb}. The present observations make
inflationary cosmology the leading theoretical framework to
explain the homogeneity, isotropy and flatness of the Universe, as
well as the observed features of the cosmic microwave background
\cite{wmap}. In most inflation models the inflaton background
dynamics for these models is usually studied in a classical
framework and in order to have a long inflationary period it is
necessary that the field rolls down very slowly: for these models
various conditions have been obtained which are different
realizations of what we will call here the {\em classical slow
roll condition}, $
 \dot{\tilde{\varphi}}^2 \ll |m^2| \; \tilde{\varphi}^2 $.
This condition guarantees that there is inflation ($ \ddot a > 0
$) and that it lasts long enough. ($ \tilde{\varphi} $ is the
classical inflaton field, $ m $ its mass, and the dot denotes
cosmic time derivative. We use the tilde, $ \tilde{\;} $, to
denote the quantities in classical inflation.)

However, since the energy scale of inflation is so high (the GUT
scale), it is necessary a full {\em quantum} field theory
description for the matter. Only such a quantum treatment permits
a consistent description of particle production and particle
decays. We address here this problem and show how the slow roll
conditions can be generalized.

\section{Quantum field inflation}  \label{quantinf}

The action for {\em quantum field inflaton} is $
  S_{q} = \tilde{S}_{g} + S_{m} + \delta\tilde{S}_{g} + \delta S_{m}
$, where $ \tilde{S}_{g} + S_{m} $ describes the dynamics of the
background, and $ \delta\tilde{S}_{g} + \delta S_{m} $ that of the
perturbations. The important difference with classical inflation
is that the dynamics of the inflaton background ($ S_m $) is
computed here in quantum field theory. The gravitational terms
have the same expressions as in the classical inflaton dynamics.
The gravitational action and its perturbation are $$
  \tilde{S}_{gr} + \delta\tilde{S}_{gr} = -\frac{1}{16\pi G}
    \int{\sqrt{-g}\;d^4\!x\,R\;}
$$, where $ G $ is the universal gravitational constant, and $ R $
is the Ricci scalar for the complete metric $ g_{\mu\nu} $. In our
treatment we consider semiclassical gravity: the geometry is
classical and the metric obeys the semiclassical Einstein
equations where the r. h. s. is the expectation value of the
quantum energy momentum tensor. (Quantum gravity corrections are
at most of order $ \sim m / M_{Planck} \sim M_{GUT} / M_{Planck}
\sim 10^{-6} $ and can be neglected.) In order to implement a
nonperturbative treatment, we consider a $N$-component inflaton
field $ \vec\chi $. The quantum matter action is $$
  S_m + \delta S_{m} = \int{\sqrt{-g}\;d^4x
    \left[\frac12\,\partial_{\alpha}\vec{\chi}\,
    \partial^{\alpha}\vec{\chi} - V(\vec{\chi}) \right]}
  = \int{d^4x\,a^3(t)\, \left[\frac12(\dot{\vec{\chi}})^2
    - \frac12\frac{(\nabla\vec{\chi})^2}{a^2(t)}
    - V(\vec{\chi}) \right]}\;,
$$ where $ \vec \chi = (\chi_1, \;\ldots, \;\chi_N) $ and $$
  V(\vec{\chi}) = \frac12\,m^2\,\vec{\chi}^2
    + \frac{\lambda}{8N}\,\left( \vec{\chi}^2 \right)^2 +
    \frac{Nm^4}{2\lambda}\frac{1-\alpha}{2}\;, $$
    with $ \alpha \equiv \mbox{sign}(m^2) = \pm 1 .
$ For positive $m^2$ the $O(N)$ symmetry is unbroken while it is
spontaneously broken for $ m^2 < 0 $. The first case describes
chaotic inflation and the second one corresponds to new inflation.
The initial state for chaotic inflation is a highly excited field
state, {\em i.e.}, a state with large $ |\tilde{\varphi}| $, while
for new inflation is a state with small $ |\tilde{\varphi}| $.

The quantum field $ \vec \chi $ can be expanded as its expectation
value $ \langle \vec\chi(x) \rangle $ plus quantum contributions
which are in general large and cannot be linearized (except for $
k/a $ much larger than the effective mass). We therefore split the
quantum contribution $ \vec\chi - \langle \vec\chi(x) \rangle $
into large quantum contributions $ \vec\varphi(x) $ (or
background), plus small quantum contributions $
\delta\vec\varphi(x) $. Thus, we express the $N$-component quantum
scalar field $\vec{\chi}$ as
\be \label{Phidec}
  \vec{\chi}(x) = \langle \vec\chi(x) \rangle + \vec\varphi(x) +
  \delta\vec\varphi(x)  \; .
\ee
After expanding the quantum matter action using
Eq.~(\ref{Phidec}), $ S_{m} $ stands for the terms without $
\delta\vec{\varphi} $ and describes the inflaton background
dynamics, while $\delta S_{m} $ stands for the remaining terms
which describe the inflaton perturbation dynamics. The dynamics of
$ \delta\vec\varphi(x) $ can be then linearized, and includes the
cosmologically relevant fluctuations, that is those which had
exited the horizon during the last $N_e \simeq 60$ efolds of
inflation. In momentum space, let us call $\Lambda $ the $k$-scale
that separates the perturbation from the background. Without loss
of generality\cite{perttsu}, we can write $ \langle \vec\chi(x)
\rangle = (\sqrt{N} \; \varphi(t),\; \vec0) \; ,$ and $
\vec{\chi}(x) = \left(\sqrt{N} \; \varphi(t) +
\varphi_{\parallel}(x), \; \vec\varphi_{\bot}(x) \right)\; + \;
\left(\delta\varphi_{\parallel}(x), \;
\delta\vec\varphi_{\bot}(x)\right) $. The mode expansions for the
fluctuations are $$
  \vec{\varphi_{\bot}}(\vec x,t) =
    \frac{1}{\sqrt{2}} \int_0^{\Lambda}
    \frac{d^3 k}{(2\pi)^3} \left[ \vec{a}_{k} \; f_{k}(t)
    \;e^{i\vec k \cdot \vec x} + \vec{a}^{\dagger}_{k} \;
    f^*_{k}(t) \; e^{-i\vec k \cdot \vec x}\right]
$$ and $$  \delta\vec\varphi_{\bot}(\vec x,t) = \frac{1}{\sqrt{2}}
    \int_{\Lambda}^\infty
    \frac{d^3 k}{(2\pi)^3} \left[\vec{a}_{k} \; f_{k}(t)
    \;e^{i\vec k \cdot \vec x} + \vec{a}^{\dagger}_{k} \;
    f^*_{k}(t) \; e^{-i\vec k \cdot \vec x}\right]
$$, with $ \vec{a}_{k} $ and $ \vec{a}^{\dagger}_{k} $ being
annihilation and creation operators, respectively, satisfying the
canonical commutation relations. [$ \varphi_{\parallel} $ and $
\delta\varphi_{\parallel} $ can also be expanded analogously with
different operators $ b_k $ and $ b^{\dagger}_k $, and modes $
g_k(t) $.] The scale $ \Lambda $ is well above the $k$-modes that
dominate the bulk of the energy, and well below the cosmologically
relevant modes. The results are independent of the precise value
of $ \Lambda $. This is due to the fact that modes with $ k \gg m
$ cannot be significantly excited since the energy density of the
universe during inflation must be of the order $ \gtrsim 10 \; m^2
\; M_{Planck}^2 $. On the other hand, modes that are relevant for
the large scale structure and the CMB are today in the range from
$ 0.1$ Mpc to $10^3$ Mpc. These scales at the beginning of
inflation correspond to physical wavenumbers in the range $
e^{N_T-60} \; 10^{16} \, GeV < k < e^{N_T-60} \; 10^{20} \, GeV $
where $N_T$ stands for the total number of efolds (see for example
Ref. \cite{sd}). Therefore, there is an intermediate $k$-range of
modes which are neither relevant for the background nor for the
observed perturbation. $\Lambda$ is inside this $k$-range, and the
results are independent of its particular value. In usual cases we
 can safely choose for $ \Lambda $,
$ 10 \; m \lesssim \Lambda \lesssim 10^3 \; e^{N_T-60} \; m \;. $

\subsection{Quantum field inflation dynamics}  \label{backqfi}

We now describe the main features of the {\em background
dynamics}, {\em i.e.}, the $ a $, $ \varphi $ and $ (
\varphi_{\parallel}, \vec{\varphi_{\bot}} ) $ dynamics. We treat
the inflaton as a full quantum field, and we study its dynamics in
a selfconsistent classical space-time metric (consistent with
inflation at a scale well below the Planck energy density). The
dynamics of the space-time metric is determined by the
semiclassical Einstein equations, where the source term is given
by the expectation value of the energy momentum tensor of the
quantum inflaton field $ G_{\mu\nu} = 8\pi m_{Pl}^{-2}\langle
T_{\mu\nu} \rangle $. Hence we solve  {\em self-consistently} the
coupled evolution equations for the classical metric and the
quantum inflaton field.

The amplitude of the quantum fluctuations for a set of modes can
be large (in quantum chaotic inflation due to the initial state,
and in new inflation due to spinodal instabilities). This implies
the need of a non-perturbative treatment of the evolution of the
quantum state, and therefore we use the large $N$ limit method. In
the large $ N $ limit, the longitudinal quantum contributions $
\varphi_{\parallel} $ are subleading by a factor $1/N$
\cite{tsuinf, tsuboya, tsunos}. Thus, the evolution equations for
the inflaton background are $$ \ddot\varphi + 3\,H\,\dot\varphi +
{\cal M}^2\,\varphi = 0, $$ and $$  \ddot f_k +3\,H\,\dot f_k +
\left(\frac{k^2}{a^2} + {\cal M}^2\right)\,f_k = 0 , \label{eqN}
$$ with $$ {\cal M}^2 = m^2 + \frac{\lambda}{2}\,\varphi^2 +
    \frac{\lambda}{2}\,\int_R{\frac{d^3k}{2(2\pi)^3}\,|f_k|^2},
$$ and for the scale factor ($ H \equiv \dot a / a $) we have $$
H^2 = \frac{8\pi}{3\,m_{Pl}^2} \; \rho $$ and $$ \frac{\rho}{N} =
\frac12 \, \dot\varphi^2 + \frac{{\cal M}^4 - m^4}{2\lambda} +
\frac{m^4}{2\lambda}\frac{1-\alpha}{2} + \frac14 \int_R
\frac{d^3k}{(2\pi)^3} \left(|\dot f_k|^2 +
\frac{k^2}{a^2}|f_k|^2\right), 
$$ where $ \rho = \langle T^{00} \rangle $ is the energy density.
The pressure ($ p\, \delta_i^{\;j} = \langle T_i^{\;j} \rangle $)
is given by $$ \frac{p+\rho}{N} = \dot\varphi^2 + \frac12 \int_R
\frac{d^3k}{(2\pi)^3} \left(|\dot f_k|^2 +
\frac{k^2}{3a^2}|f_k|^2\right)\;. 
$$ The index $ R $ denotes the renormalized expressions of these
integrals \cite{perttsu,tsuinf}. This means that we must subtract
the appropriate asymptotic ultraviolet behaviour in order to make
convergent the integrals. The evolution equations for the
expectation value and for the field modes are analogous to damped
oscillator equations, and the inflationary period ($ \ddot a > 0
$) corresponds to the overdamped regime of these damped
oscillators.

We consider here two typical classes of quantum inflation models:
\begin{itemize}
\item
{\em (i) Quantum chaotic inflation}, where inflation is produced
by the dynamical {\em quantum} evolution of a excited initial pure
state with large energy density (more details and the
generalization to mixed states can be found in \cite{tsuinf}).
This state is formed by a distribution of excited modes. It can be
shown that the initial conditions for a general pure state are
given by fixing the complex values of $ f_k(0) $ and $ \dot f_k(0)
$. Among these four real (two complex) numbers for each $ k $
mode, one is an arbitrary global phase, and another is fixed by
the wronskian. The two remaining degrees of freedom fix the
occupation number for each mode and the relative phase between $
f_k(0) $ and $ \dot f_k(0) $. The coherence between different $ k
$ modes turns out to be determined by such relative phases.

\item
{\em (ii) Quantum new inflation}, where inflation is produced by
the dynamical {\em quantum} evolution of a state with small
inflaton expectation value, and small occupation numbers for the
quantum modes, evolving with a spontaneously broken symmetry
potential. (More details can be found in \cite{qnew}.)

\end{itemize}

The two classes of quantum inflation models have important
differences in their initial state and in their background and
perturbation dynamics ({\em e.g.}, spinodal instabilities are
present in new inflation and not in chaotic inflation). However,
we stress here the common features which allow a unified treatment
of the computation for the primordial perturbations generated in
these models. In this quantum field inflation framework we have
found the following \emph{generalized slow roll condition}
\be \label{gsrc}
\dot\varphi^2+\int_R{\frac{d^3k}{2(2\pi)^3}\,|\dot f_k|^2} \;\ll\;
m^2 \left( \varphi^2 + \int_R{\frac{d^3k}{2(2\pi)^3}\,|f_k|^2}
\right)
\ee
which guarantees inflation ($\ddot a > 0$) and that it lasts long
(for both scenarios). (This condition includes the classical slow
roll condition $ \dot\varphi^2 \ll m^2 \; \varphi^2 $ as a
particular case.) There is a wide class of quantum initial
conditions satisfying Eq.(\ref{gsrc}) and leading to inflation
that lasts long enough \cite{tsuinf}. The quantum field dynamics
considered here leads to \emph{two inflationary epochs}, separated
by a condensate formation:

\begin{enumerate}

\item \emph{The pre-condensate epoch:} During this epoch the term
$$ D \equiv \int \frac{d^3k}{(2\pi)^3} \; \frac{k^2}{a^2} \;
|f_k|^2 $$ has an important contribution to the energy density
while it fastly decreases due to the exponential redshift of the
excitations ($ k/a \to 0 $). This epoch ends at a time $\tau_A $
when the $ D $ contribution to the energy density becomes
negligible, {\em i.e.}, the $ k^2/a^2 $ contribution in the
background evolution equations is negligible at $ \tau = \tau_A $.

After outward horizon crossing, the time dependence of the modes
factorizes and becomes $ k $ independent. The $ k^2/a^2 $ term in
the mode evolution equations becomes negligible, and all the modes
satisfy the same damped oscillator equation. For $ m^2 > 0 $ the
modes decrease (due to the damping), while for $ m^2 < 0 $ they
grow (due to spinodal instabilities). At the end of this epoch ($
t = \tau_A $) all the relevant modes for the background dynamics
have exited the horizon, and the time dependence factorization
allows to consider them as a zero mode condensate.

\item \emph{The post-condensate  quasi-de Sitter epoch}.
The enormous redshift of the previous epoch assembles the quanta
into a zero mode condensate, $ \tilde{\varphi}_{eff} $, given
by\cite{qnew} $ \tilde{\varphi}_{eff}^1(t) = \sqrt{N} \;
\varphi(t) $, and $ \tilde{\varphi}_{eff}^i(t) = \sqrt{
\int{\frac{d^3k}{2(2\pi)^3}\,|f_k(t)|^2}}$ for $ i = 2, \ldots, N
$ with constant direction in the field space [due to the $ O(N) $
invariance of the potential], and modulus $$
\tilde{\varphi}_{eff}(t) = \sqrt{N}
\sqrt{\varphi^2(t)+\int{\frac{d^3k}{2(2\pi)^3}\,|f_k(t)|^2}} $$
that verifies the classical equations of motion, $$
\ddot{\tilde{\varphi}}_{eff} + 3 \,H \,\dot{\tilde{\varphi}}_{eff}
  + \tilde m^2 \; \tilde{\varphi}_{eff} + \tilde{\lambda}\;
  \tilde{\varphi}_{eff}^3 = 0 \label{eqclasphi} $$ and
$$ H^2 = \frac{8 \pi}{3 m_{Pl}^2} \; \rho $$ where $$
 \rho = \frac12\; \dot{\tilde{\varphi}}_{eff}^2
 + \frac12\; \tilde m^2 \; \tilde{\varphi}_{eff}^2 +
 \frac{\tilde{\lambda}}{4}\;\tilde{\varphi}_{eff}^4 $$
with $ \tilde{\lambda} = \frac{\lambda}{2 N} $, $ \tilde m^2 = m^2
$, $ \tilde \alpha \equiv \mbox{sign}(\tilde m^2) = \pm 1 $. The
pressure is given by $ p + \rho = \dot{\tilde{\varphi}}_{eff}^2 $.
Therefore, the background evolution in this period can be {\em
effectively} described by a {\em classical} scalar field obeying
the evolution equation (\ref{eqclasphi}) and with initial
conditions defined at $ t = \tau_A $. Moreover, it is important to
stress that the {\em initial conditions for $
\tilde{\varphi}_{eff} $ are fixed by the quantum state}: $$
\tilde{\varphi}_{eff}(\tau_A) = \sqrt{N}
\sqrt{\varphi^2(\tau_A)+\int{\frac{d^3k}{2(2\pi)^3}\,|f_k(\tau_A)|^2}}\;.
$$ Also the value of $ \tau_A $ depends on the full quantum
evolution {\em before} the formation of the condensate. $ \tau_A $
is therefore a function of the coupling, the mass and the quantum
initial conditions \cite{tsuinf}.
\end{enumerate}

The previous result shows that after the formation of the
condensate (both for chaotic and for new inflation), the
background dynamics can be described by an effective classical
background inflation whose action structure, parameters (mass and
coupling) and initial conditions are fixed by those of the
underlying quantum field inflation.

In chaotic inflation the total number of efolds $N_T$ {\em
decreases} if the initial state had excited modes with non-zero
wavenumber (for constant initial energy). For example, if the
initial energy is concentrated in a shell of wavenumber $ k_0 $
and for simplicity the quadratic term dominates the potential
\cite{tsuinf}, we have $ N_T \simeq \frac{4\pi}{m_{Pl}^2\,m^2}\;
\frac{\rho_0}{\;1+(k_0/m)^2\;} $ (where the classical result is
recovered at $ k_0 = 0 $). We have shown \cite{tsuinf} that there
are enough efolds even for $k_0 \sim 80 \, m $ for reasonable
choices of the initial energy density ($ \rho_0 = 10^{-2} m_{Pl}^4
$) and of the parameters (for instance, $ N \; m^2 /[ \lambda \;
m_{Pl}^2 ] = 2 \cdot 10^5 $).

\subsection{Primordial perturbations in quantum field inflation}
\label{pertqfi}

The relevant primordial perturbations are those that exited the
horizon during the last efolds of inflation \cite{infbooks}. As we
have seen in the previous subsection the background ($ a $, $
\varphi $, $ \varphi_{\parallel} $ and $ \vec{\varphi_{\bot}} $)
dynamics during the last efolds in the quantum field inflation
scenarios (both chaotic and new) are effectively classical. This
will allow to compute the relevant primordial perturbations for
these scenarios and express them in terms of the known
perturbations for the corresponding single-field classical
scenarios.

The  more general metric perturbation $\delta g_{\mu\nu} $ can be
decomposed as usual in scalar, vector and tensor components
\cite{perttsu, revmuk, multiftent}. In the linear approximation,
scalar, vector and tensor perturbations evolve independently and
thus can be considered separately \cite{revmuk}.

\paragraph{Scalar perturbations}

We now compute the {\em scalar metric perturbations} to the
background, these are tightly coupled to the inflaton
perturbations, and therefore, both perturbations have to be
studied together \cite{revmuk,revriotto}. We have shown that the
background dynamics for quantum field inflation can be separated
in two epochs: before and after the formation of the condensate.
As the first one is short, in the more natural scenarios the last
$ N_e \simeq 60 $ efolds take place after the formation of the
condensate. Thus, the cosmologically relevant scales of the
perturbations exit the horizon when the condensate was already
formed. Therefore, the dynamics of the perturbations after the
formation of the condensate is well approximated by that given by
the effective classical inflation background. In addition the $
O(N) $ invariance of the potential implies that the isocurvature
scalar perturbations are negligible, and that the background
inflaton has a straight line trajectory in field space giving for
the power spectrum of {\em adiabatic scalar perturbations} in the
quantum field inflation\cite{perttsu} the result $$ |
\delta_{k\;ad}^{(S)}(m^2,\lambda)|^2 = |
\tilde\delta_{k\;ad}^{(S)}(m^2,\;\frac{\lambda}{2 N} )
   |^2
$$ where $ |\tilde\delta_k^{(S)}(m^2,\tilde{\lambda})|^2 $ is the
power spectrum of scalar perturbations for the {\em single-field}
classical background inflation (evolution equations and initial
conditions of the post-condensate regime). The corrections due to
the next to leading order in large $N$ are estimated to be of the
order of $ \frac{m^2}{N \; H^2} $, while the correction due to
non-vacuum initial conditions caused by the dynamics previous to
the condensate formation are estimated to be of the order of $
\frac{m^2}{H^2} $, where $m$ is the inflaton mass and $H$ the
Hubble constant at the moment of horizon crossing. An upper
estimate gives about $ 4\% $ for each of them in the
cosmologically relevant scales.

\paragraph{Vector perturbations}

The vector metric perturbations do not have any source in their
evolution equation, because the energy-momentum tensor for a
scalar field does not lead to any vector perturbation. The $ (0i)
$ components of the Einstein equations, in the absence of vector
perturbation sources, gives $ \Delta S_i = 0 \;, $ implying that
there cannot be any space-dependent vector perturbations [in
Fourier space $ k^2 S_i(k) = 0 $]. Therefore, the \emph{vector
perturbations} are \emph{negligible} (as for classically driven
inflation \cite{multiftent}). $ |\delta_k^{(V)}(m^2, \lambda)|^2 =
0 \;. $

\paragraph{Tensor perturbations}

As the energy-momentum tensor of a scalar field do not have tensor
perturbations, the tensor metric perturbations do not have any
source in their equation. Therefore, the amplitude of tensor
perturbations is determined only by the background evolution,
which after the condensate formation has {\it an effective
single-field classical description}. Thus, the tensor
perturbations for the quantum inflation scenario are $$
|\delta_k^{(T)}(m^2,\lambda)|^2 =
    | \tilde\delta_k^{(T)}(m^2,\;\frac{\lambda}{2 N} )
   |^2
$$ where $ |\tilde\delta_k^{(T)}(\tilde m^2,\tilde{\lambda})|^2 $
is the power spectrum of tensor perturbations for {\it
single}-field classical inflation.

\paragraph{Tensor to scalar amplitude ratio}

The tensor to scalar ratio $ r $ is defined as $ r \equiv
|\delta_{k}^{(T)}|^2 / |\delta_{k\;ad}^{(S)}|^2 \;. $ From the
previous expressions for the spectra of tensor and adiabatic
scalar perturbations, it follows that $$ r(m^2,\lambda) =  \tilde
r \left(m^2,\;\frac{\lambda}{2 N} \right)$$ where $ \tilde r
(m^2,\lambda) $ is the tensor to scalar amplitude ratio for
single-field classical inflation.

\section{Conclusions}  \label{conclu}

Quantum field inflation (under conditions that are the quantum
analog of slow roll) leads, upon evolution, to the formation of a
condensate starting a regime of effective classical single-field
inflation. The action structure, parameters (mass and coupling)
and initial conditions for the effective classical field
description are fixed by those of the underlying quantum field
inflation. We show that this effective description allows an easy
computation of the primordial perturbations which takes into
account the dominant quantum effects (quantum inflaton background
and quantum inflaton and metric perturbations), and gives the
results in terms of the classical inflation results. In
particular, isocurvature scalar perturbations are absent (at first
order of slow roll) due to the $ O(N) $ invariance of the
potential  in agreement with the WMAP data. It is thus the
presence of a large symmetry in multi field models that make them
compatible with the present observations.

For chaotic inflation, the quantum treatment avoids the unnatural
requirements of an initial state with all the energy in the zero
mode. For new inflation it allows a consistent treatment of the
explosive particle production due to spinodal instabilities.
Quantum field inflation provides enough efolds of inflation
provided the generalized slow roll condition is fulfilled. In the
chaotic quantum field inflation the number of efolds is lower than
in classical inflation when modes with non-zero wavenumber are
excited initially. Quantum corrections to the scalar perturbations
power spectrum due to initial conditions and from next to leading
order in large $N$ are estimated to be at most of the other of $ 4
\% $ each of them.

In summary, the classical inflationary scenario emerges as an
effective description of the post-condensate inflationary period
both for the {\em background} and for the {\em perturbations}.
Therefore, this generalized  inflation provides the {\em quantum
field foundations} for classical inflation, which is in agreement
with CMB anisotropy observations \cite{wmap}.

\ack
F. J. C. thanks N. G. Sanchez, H. J. de Vega and D. Boyanovsky for
the pleasant collaborations that has lead to these results. F. J.
C. acknowledges support from MECD (Spain) through research
projects BFM2003-02547/FISI and FIS2006-05895.

\end{document}